\documentclass[11pt,twoside,a4paper]{cernyrep}

\usepackage{lineno}
\usepackage{graphicx}%
\usepackage{multirow}%
\usepackage{amsmath,amssymb,amsfonts}%
\usepackage{amsthm}%
\usepackage{mathrsfs}%
\usepackage[title]{appendix}%
\usepackage[table,xcdraw]{xcolor}
\usepackage{textcomp}%
\usepackage{manyfoot}%
\usepackage{booktabs}%
\usepackage{algorithm}%
\usepackage{algorithmicx}%
\usepackage{algpseudocode}%
\usepackage{listings}%
\usepackage{placeins}
\usepackage{subfig}
\usepackage{subfiles}%
\usepackage{textgreek}
\usepackage{comment}
\usepackage{siunitx}
\usepackage{tikz} 
\usepackage{tabularx}       
\usepackage{makecell}       
\usepackage{pdflscape} 
\usepackage{booktabs} 
\usepackage{amsmath} 
\usepackage{makecell}
\usepackage{multirow} 
\usepackage{tcolorbox} 
\usepackage[normalem]{ulem}
\usepackage{cancel}
\usepackage{wrapfig}
\usepackage{rotating}
\usepackage{wasysym}   
\usepackage{hyperref}
\usepackage{enumitem}
\hypersetup{%
        colorlinks,
        breaklinks=true,
        plainpages=false,%
        citecolor=blue,
        linkcolor=blue,
        urlcolor=blue,
        bookmarksopen=true,%
        bookmarksnumbered=false,%
        bookmarksdepth=5%
}

\newcommand{\micron}{\ensuremath{\upmu \mathrm{m}}}

\newcommand{\keV}{\si{\kilo\electronvolt}}
\newcommand{\MeV}{\si{\mega\electronvolt}}
\newcommand{\GeV}{\si{\giga\electronvolt}}

\newcommand{\mlab}[1]%
    {\mbox{}\marginpar{\raggedright\hspace{0pt}\tiny {\color{red}{#1}}}}

\newcommand{\as}{\alpha_{\rm S}}

\newcommand{\invab}{{\rm ab}^{-1}}

\providecommand{\ttbar}{\PQt\PAQt}

\begin{document}

\pagestyle{empty}

\begin{center}

\textbf{\Large The FCC integrated programme: a physics manifesto }
\vspace{4cm}

Editors:\\
A.~Blondel\,$^{1}$,
C.~Grojean\,$^{2}$,
P.~Janot\,$^{3}$,
S.~Rajagopalan,$^{4}$,
G.~Wilkinson\,$^{3,5}$ \\
\vspace{0.3cm}

\noindent
$^{1}$ LPNHE-CNRS/IN2P3, Sorbonne University, Paris, France \\ and DPNC, University of
Geneva, Switzerland, \\
$^{2}$ Deutsches Elektronen-Synchrotron DESY, Hamburg and Humbolt-Universit\"{a}t zu Berlin, Germany,\\
$^{3}$ CERN, Switzerland, 
$^{4}$ Brookhaven Laboratory, USA,
$^{5}$ University of Oxford, UK
\vspace*{0.5cm}

on behalf of the Physics, Experiments and Detectors pillar of the FCC Project, \\ including those who signed the FCC Feasibility Study Report, and those who \\ contributed  to the FCC documents referenced herein

\end{center}

\vspace*{0.5cm}
\begin{center}
{\bf Abstract}
\end{center}

The FCC integrated programme comprises an $\rm e^+e^-$ high-luminosity circular collider that will produce very large samples of data in an energy range $88 \le \sqrt{s} \le 365$\,GeV, followed by a high-energy $\rm pp$ machine that, with the current baseline plan, will operate at a collision energy of around 85\,TeV and deliver datasets an order of magnitude larger than those of the  HL-LHC. 
This visionary project will allow for transformative measurements across a very broad range of topics, which in almost all cases will exceed in sensitivity the projections of any other proposed facility, and simultaneously provide the best possible opportunity for discovering physics beyond the Standard Model. The highlights of the physics programme are presented, together with discussion on the key attributes of the integrated project that enable the physics reach.  It is noted that the baseline programme of FCC-ee, in particular, is both flexible and extendable, and also that the synergy and complementarity of the electron and proton  machines, and the sharing of a common infrastructure, provides a remarkably efficient, timely and cost-effective approach to addressing the most pressing open questions in elementary particle physics.


\vfill
\begin{center}
\emph{\today} \\
\vspace{1cm}
Contribution to the European Strategy for Particle Physics Update 2025-2026
\end{center}

\newpage

\pagenumbering{arabic}
\setcounter{page}{1}


\section{Physics context}
\vspace*{-0.1cm}

The particle physics landscape has been profoundly influenced by the discovery of a Higgs boson with a mass around 125\,GeV at the LHC~\cite{CMS:2012qbp,ATLAS:2012yve}, which completes the observation of all the particles predicted by the Standard Model (SM).
This consistent and predictive theory, for which all the free parameters are now measured, has so far been successful at describing all phenomena accessible to collider experiments.
After almost fifteen years of LHC operation, both precision measurements and many exploratory searches have demonstrated the validity of the SM and excluded signs of New Physics across a wide range of energies around the TeV scale.

This notwithstanding, core parts of the Higgs sector of the SM still remain to be confronted experimentally, and several fundamental empirical facts remain unexplained, such as the abundance of matter over antimatter, the evidence for dark matter, and  the non-zero value of neutrino masses. These unanswered questions, and several theoretical puzzles, such as the low value of the Higgs mass, the three generations of matter, and the number of free parameters in the theory, point to the existence of a more complete, but as yet unknown, theory of physics beyond the  Standard Model (BSM), and mandates a new programme of collider exploration after the completion of the HL-LHC.  In contrast to previous decades, however, the signposts provided by theory do not point in one single direction.   Solutions to the shortcomings of the SM may exist at even higher energy, at the price of either an \emph{unnatural} value of the weak scale or an ingenious, but still elusive structure. There are also strong motivations for the existence of light and feebly coupled new particles~\cite{Graham:2015cka, Espinosa:2015eda, Arkani-Hamed:2020yna, Asaka:2005pn}.  Neither the mass scale (from meV to ZeV) of this New Physics nor the strength of the couplings to the SM (from 1 to $10^{-12}$ or less) are known, thus calling for a discovery programme with the highest sensitivity across the broadest range of scenarios, exploring both the intensity and the energy frontiers.

The 2020 update of the European Strategy for Particle Physics (ESPP) recommended that ``Europe, together with its international partners, should investigate the technical feasibility of a future hadron collider at CERN with a centre-of-mass energy of at least 100\,TeV and with an electron-positron Higgs and electroweak factory as a possible first stage''~\cite{esppu}.  This statement consolidated a similar recommendation from the 2013 ESPP Update~\cite{espp2013}, giving additional emphasis to the importance of an $\rm e^+e^-$ machine for precise studies of the Higgs, electroweak and flavour sectors, and recognising the growing realisation in the community that the physics landscape being revealed by the LHC favours a high-luminosity circular collider as the natural facility at which to pursue these studies. Following the 2020 ESPP Update, the CERN Council launched the FCC Feasibility Study, which has recently delivered its Final Report (FSR) on the integrated programme of the $\rm e^+e^-$ (FCC-ee) and hadron (FCC-hh) machines~\cite{FCC-FSR-Vol1,FCC-FSR-Vol2,FCC-FSR-Vol3}.  This study builds on an earlier phase of investigations, documented in the Conceptual Design Report (CDR)~\cite{fcc-phys-cdr,fcc-ee-cdr,FCC-hhCDR}.  
The FCC integrated programme is also in full accordance with the 2023 US P5 recommendations, which prioritised ``an offshore Higgs factory'' and an R\&D path towards a ``10\,TeV pCM [parton centre-of-mass] collider''~\cite{usp5}.  

The FCC project is fully aligned with the priorities of the international community as summarised above, and the needs of physics  at this critical juncture.  Its science programme exceeds those of all alternative options in both breadth and depth, and is optimal in terms of its environmental impact~\cite{Blondel:2024mry}.
Its goals are to:
\vspace*{-0.2cm}
\begin{itemize}[leftmargin=0.4cm,itemsep=0mm, parsep=0pt]
\item Map the properties of the Higgs and electroweak gauge bosons, with accuracies order(s) of magnitude better than will be known at the end of the HL-LHC;
\item Further pursue the quest to determine the origins of the masses of the quarks and leptons, and conclusively establish or refute the central pillar of the SM, its hypothesis for the Higgs potential, acquiring sensitivity to the processes that led to the formation of today's Higgs vacuum field during the period $10^{-12}$ and $10^{-10}$ seconds after the Big Bang;
\item Sharpen significantly our knowledge of already explored particle-physics phenomena with a comprehensive and accurate campaign of highly precise  measurements sensitive to the effect of mass scales far beyond the direct kinematic reach;
\item Improve by orders of magnitude the sensitivity to rare and elusive phenomena at low energies, including the possible discovery of light particles with very small couplings, e.g. massive neutrinos or axion-like particles;
\item Improve by an order of magnitude the direct discovery reach for new particles at the energy frontier.
\end{itemize}
This submission to the 2025-2026 ESPP Update presents an overview of the science programme of FCC, summarises its most important deliverables, and highlights the critical aspects of the project that will allow these goals to be attained. Accompanying documents provide more information on FCC-ee and FCC-hh physics in the domains of Higgs, electroweak and top measurements~\cite{EPPSU-FCC-HEWT}, QCD studies~\cite{EPPSU-FCC-QCD}, flavour physics~\cite{EPPSU-FCC-flavour} and searches for BSM physics~\cite{EPPSU-FCC-BSM}.  Further discussion on physics considerations specific to FCC-hh can be found in Ref.~\cite{EPPSU-FCC-hhspecifities}.  Summaries of the FCC-ee and FCC-hh projects as a whole are provided in Refs.~\cite{EPPSU-FCC-FCCee, EPPSU-FCC-FCChh}.  Full information on the FCC integrated programme is available in the FSR~\cite{FCC-FSR-Vol1,FCC-FSR-Vol2,FCC-FSR-Vol3}.  Appendix~\ref{sec:organisation} contains information on the organisation of FCC physics, experiments and detector activities.

\vspace*{-0.5cm}
\section{FCC physics: guiding principles and selected highlights} 
\vspace*{-0.1cm}

The physics programme of FCC, in both its $\rm e^+e^-$ and $\rm pp$ phases, is immensely broad and can deliver  transformational measurements over a wide span of topics. A  select summary is provided below, with more information being available in Volume~1 of the FSR~\cite{FCC-FSR-Vol1}, and in dedicated documents on specific physics areas that accompany this submission~\cite{EPPSU-FCC-HEWT,EPPSU-FCC-QCD,EPPSU-FCC-flavour,EPPSU-FCC-BSM,EPPSU-FCC-hhspecifities} .  

The measurements at FCC-ee will originate from datasets taken at energies optimised for the study of all the heavy particles of the Standard Model, with the baseline target samples summarised in Table~\ref{tab:seqbaseline}.
The properties of the Higgs will be studied through large datasets collected at 240\,GeV and 340-365\,GeV, allowing for both the Higgs-strahlung and $\PW\PW$-fusion processes to be exploited, with the higher energies also serving as a laboratory for top physics.
Of particular note is the so-called Tera-Z run, which will result in a sample of $6 \times 10^{12}$ \PZ-boson decays.  The \PW mass and width will be measured in a run at and around the $\PW\PW$ threshold.  The hadron collider,
FCC-hh, will deliver $\rm pp$ collisions at energies and luminosities far in excess of the HL-LHC.  The current set of parameters assumed in the FSR~\cite{FCC-FSR-Vol1,FCC-FSR-Vol2,FCC-FSR-Vol3} corresponds to a collision energy of 84.6\,TeV and an integrated luminosity of 30\,${\rm ab^{-1}}$, sufficient for the production of $\sim$20~billion Higgs bosons and allowing for the direct discovery of new particles with masses up to several tens of TeV. If suitable technological choices are made in the design of the collider and its injectors, FCC-hh will also be able to accommodate ion collisions at collision energies and luminosities that are an order of magnitude above those of the HL-LHC.  Although not part of the baseline integrated programme, it would also be possible to extend the FCC complex to make provision for electron-proton and electron-ion  collisions.

\begin{table}[ht]
\renewcommand{\arraystretch}{0.95}
\centering
\caption{\small The baseline FCC-ee operation model with four interaction points, showing the centre-of-mass energies, 
instantaneous luminosities for each IP, and integrated luminosity per year summed over four~IPs.  
The integrated luminosity values correspond to 185 days of physics per year and a 75\% operational efficiency (i.e., $1.2 \times 10^7$ seconds per year)~\cite{Bordry:2645151}.
The last two rows indicate the total integrated luminosity and number of events expected to be produced in the four detectors. 
The number of $\PW\PW$ events includes all $\sqrt{s}$ values from 157.5\,GeV up.
\label{tab:seqbaseline}}
\begin{tabular}{lccccccc}
\hline 
Working point & \PZ pole & $\PW\PW$ thresh.\ & $\PZ\PH$ & \multicolumn{2}{c}{$\PQt\PAQt$} \\ \hline
$\sqrt{s}$ {(GeV)} & 88, 91, 94 & 157, 163 & 240 & 340--350 & 365 \\ 
Lumi/IP {($10^{34}$\,cm$^{-2}$s$^{-1}$)} & 140 & 20 & 7.5 & 1.8 & 1.4 \\ 
Lumi/year {(ab$^{-1}$)} & 68 & 9.6 & 3.6 & 0.83 & 0.67 \\ 
Run time {(year)} & 4 & 2 & 3 & 1 & 4 \\ 
Integrated lumi.\ {(ab$^{-1}$)} & 205 & 19.2 & 10.8 & 0.42 & 2.70 \\ \hline
 &  &  & $2.2 \times 10^6$ $\PZ\PH$ & \multicolumn{2}{c}{$2 \times 10^6$ $\PQt\PAQt$} \\
Number of events &  $6 \times 10^{12}$ \PZ & $2.4 \times 10^8$ $\PW\PW$ & $+$ & \multicolumn{2}{c}{$+\,370$k $\PZ\PH$} \\
 &  &  & 65k $\PW\PW \to \PH$ & \multicolumn{2}{c}{$+\,92$k $\PW\PW \to \PH$} \\ \hline
\end{tabular} 
\end{table}

The following summary makes clear the exceptional reach that FCC will have in direct searches for BSM physics.  Accompanying this, however, is a measurement campaign of extreme and unprecedented precision, encompassing the Higgs sector, electroweak studies and flavour physics.  It is important to appreciate that these measurements themselves have powerful BSM search potential. To give one example, the precision of the measurements that will be performed in the Tera-Z run will be sensitive, in a very general manner, to the impact of New Physics.
Indeed,  the electroweak observables measured at FCC-ee, together with the sub-percent determination of the Higgs couplings, will provide an {\it almost inescapable probe} for any New Physics model that involves particles of ${\cal{O}}$(TeVs) mass and contributes at tree-level to SM modifications and, for many models, give sensitivity up to several tens of TeV, an order of magnitude improvement on LEP/HL-LHC~\cite{Allwicher:2024sso,Gargalionis:2024jaw,Maura:2024zxz,terHoeve:2025gey}. (The one loophole to this assertion - that the New Physics may be feebly interacting - is itself addressed by the direct-search programme of FCC-ee.) It is unlikely that any deviation with respect to the SM expectation will appear in isolation, and so the very wide expanse of FCC-ee measurements will be critical in helping elucidate the nature of the BSM physics that might manifest itself.   However, the FCC-hh will then be essential in giving the possibility of producing these new heavy particles on-shell, and studying their properties in detail.  Furthermore, the FCC-hh will be able to extend the direct window to still higher mass scales, and to perform complementary indirect searches through its own programme of precision measurements.

\vspace*{-0.5cm}
\subsection*{Higgs physics}

The FCC will perform a comprehensive and highly precise characterisation of the properties of the Higgs boson.  Expectations for several of the key measurements are presented in Table~\ref{tab:HiggsKappa3}. 

At FCC-ee, it will be possible to measure the Higgs width directly, and therefore provide model-independent measurements of the branching fractions, which is not possible at the HL-LHC.   Many of these branching fractions will be measured with sub-percent uncertainties, which is critical for providing sensitivity to BSM effects, and with a precision significantly better than would be achievable at any alternative $\rm e^+e^-$-collider project proposed for CERN.  FCC-ee will also provide the best opportunity for measuring the strange-Higgs coupling and is the only facility that offers the  possibility to achieve sensitivity to the electron-Higgs coupling at around the SM level, thereby, for the first time, exploring the interaction of the Higgs field with the building blocks of everyday matter.   The measurement of the single-Higgs production cross sections at 240\,GeV and 365\,GeV will give access to the Higgs self-coupling through loop-level processes, often expressed through the parameter $\kappa_\lambda$, in a manner complementary to the di-Higgs production measurements at HL-LHC. The precision on $\kappa_\lambda$ from the combination of HL-LHC and the FCC-ee results is expected to reach 18\% or better, and can be further improved through modifications to the baseline FCC-ee run plan.  

\begin{table}[ht]
\centering
\caption{\small Expected 68\%~CL relative precision of the $\kappa$ parameters (Higgs couplings relative to the SM) 
and of the Higgs boson total decay width $\Gamma_{\PH}$, 
together with the corresponding 95\%~CL upper limits on the untagged (undetected events), $\cal{B}_{\text{unt}}$, 
and invisible, $\cal{B}_{\text{inv}}$, branching ratios at HL-LHC, FCC-ee (combined with HL-LHC), and the FCC integrated programme.
For the HL-LHC numbers, a $|\kappa_V| \leq 1$ constraint is applied (denoted with an asterisk), 
since no direct access to $\Gamma_{\PH}$
is possible;
this restriction is lifted in the combination with FCC-ee.
The `--' indicates that a particular parameter has been fixed to the SM value, due to lack of sensitivity.
From Ref.~\cite{deBlas:2019rxi}, updated with four~IPs, the baseline luminosities of Table~\ref{tab:seqbaseline}, and the most recent versions of the analysis. 
For some of the entries, the $\kappa$ precision starts being limited by the projected SM parametric uncertainties, e.g. in $m_{\PQb}$~\cite{Freitas:2019bre}. In these cases, the result obtained by neglecting such parametric uncertainties is also reported (separated by a $/$).
}
\label{tab:HiggsKappa3}
\begin{tabular}{ c   c    c     c }
\toprule
Coupling & HL-LHC 
& FCC-ee 
& FCC-ee $+$ FCC-hh\\ 
\midrule 
$\kappa_{\PZ}$ (\%) &  1.3$^\ast$ 
& 0.10  
& 0.10 \\ 
$\kappa_{\PW}$ (\%)  &   1.5$^\ast$  
& 0.29 
& 0.25 \\
$\kappa_{\PQb}$ (\%)  & 2.5$^\ast$  
& 0.38 / 0.49 
& 0.33 / 0.45 \\ 
$\kappa_{\Pg}$ (\%)    &   2$^\ast$ 
& 0.49 / 0.54 
& 0.41 / 0.44 \\ 
$\kappa_{\PGt}$ (\%)   & 1.6$^\ast$    
& 0.46 
& 0.40 \\
$\kappa_{\PQc}$ (\%)   & --   
&0.70 / 0.87 
& 0.68 / 0.85 \\
$\kappa_{\PGg}$ (\%)  &  1.6$^\ast$  
& 1.1 
& 0.30 \\
$\kappa_{\PZ\PGg}$ (\%)    & 10$^\ast$  
& 4.3
& 0.67\\
$\kappa_{\PQt}$ (\%)   & 3.2$^\ast$ 
& 3.1 
& 0.75 \\
$\kappa_{\PGm}$ (\%)     & 4.4$^\ast$  
& 3.3 
& 0.42 \\
$\vert \kappa_{\PQs}\vert$ (\%) & -- & $_{-67}^{+29}$ & $_{-67}^{+29}$\\ 
$\Gamma_{\PH}$ (\%) & -- 
& 0.78
& 0.69 \\ 
$\cal{B}_{\text{inv}}$ ($<$, 95\% CL) 
& $1.9 \times 10^{-2}$ $^\ast$   
& $5 \times 10^{-4}$ 
& $2.3 \times 10^{-4}$ \\
$\cal{B}_{\text{unt}}$ ($<$, 95\% CL) & $4 \times 10^{-2}$ $^\ast$ 
& $6.8\times 10^{-3}$ 
& $6.7\times 10^{-3}$\\
\bottomrule
\end{tabular}
\end{table}

The FCC-hh will complete this programme, by improving the knowledge of these couplings still further, in particular $\rm Ht\bar{t}$ and those couplings to  suppressed final states such as $\rm H\gamma \gamma$, $\rm H\mu^+\mu^-$ and $\rm HZ \gamma$.  Most importantly, the Higgs potential will be tightly constrained through a measurement of $\kappa_\lambda$ with percent-level precision.

\vspace*{-0.5cm}
\subsection*{Electroweak and top physics}

The enormous expected improvement in the knowledge of electroweak observables is one of the most remarkable features of the FCC-ee programme, and may  turn out to hold the greatest physics significance.

The Tera-Z run will have immense statistical power. Strategies are already identified that will allow for the corresponding systematic control that will be required, for example full exploitation of the excellent beam-energy calibration that resonant depolarisation  at storage rings uniquely allows.  To give but three examples, it will be possible to improve the precision of the Z peak hadronic cross section $\sigma^0_{\rm had}$ by a factor 40, that of $\sin^2\theta^{\rm eff}_{\rm W}$ by a factor 90 (from the forward-backward dimuon asymmetry alone), and that of the Z-width itself by a factor of almost 200.  
These are all current estimates,  and further studies may point the way to still higher precision. In many cases theoretical work, now underway, will also be required to match the experimental sensitivity~\cite{FCC-FSR-Vol1}.
Other measurements will be possible for the first time, such as a direct measurement of $\alpha_{\rm QED}(m_{\rm Z}^2)$ through interference effects in off-peak forward-backward asymmetries~\cite{Janot:2015gjr}, or through studies of differential distributions on-peak~\cite{Riembau:2025ppc}.

\begin{figure}[t]
\begin{center}
\includegraphics[width=0.5\linewidth]{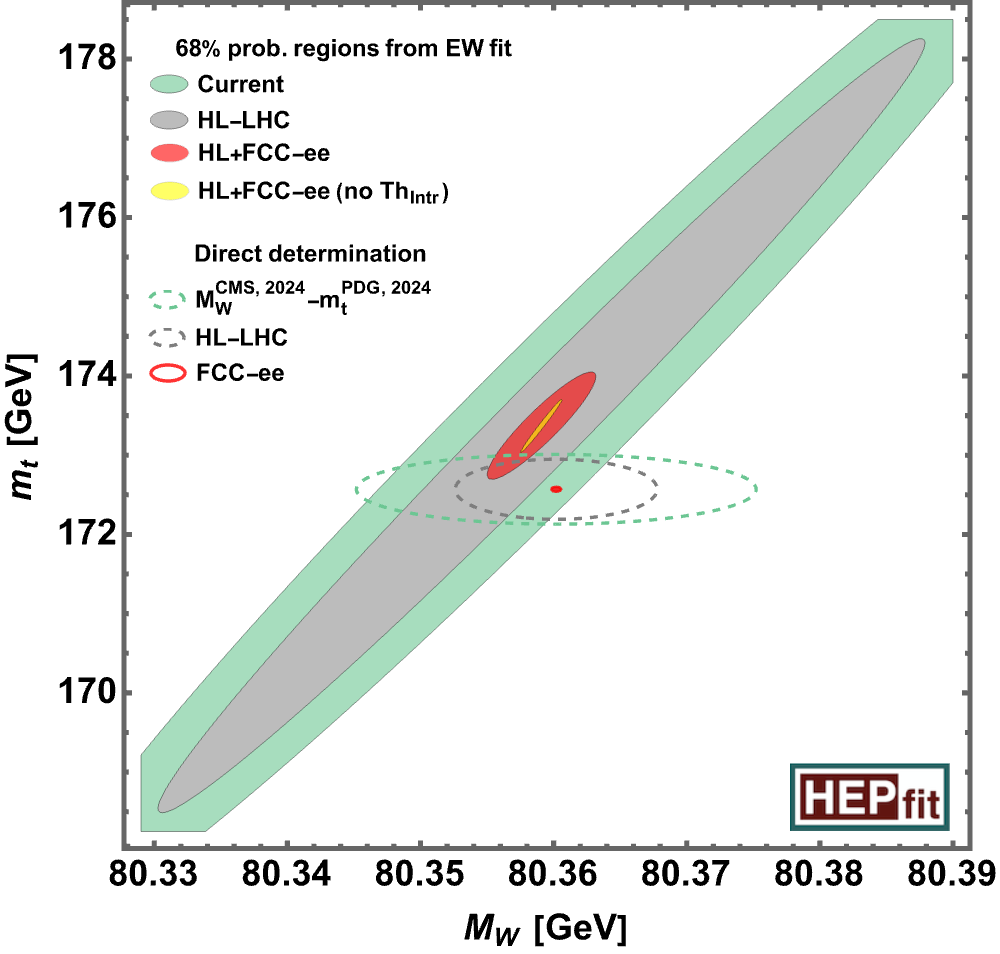}
\end{center}
\vspace*{-0.3cm}
\caption{\small Direct measurements and expectations from the electroweak fit of $\rm m_W$ and $m_t$, showing current status, expected status at the end of HL-LHC, and expected status after FCC-ee.  All projections are centred around the current central values~\cite{Jorge}.  
}
\label{fig:mwvsmt}
\end{figure}

The very large event samples and precise knowledge of the beam energy will also be decisive factors at and around the $\PW\PW$ threshold, allowing FCC-ee to measure ${\rm m_W}$ with an uncertainty of around 230\,keV, which is a factor 50 better than the world's current best measurement~\cite{CMS:2024lrd}.  
This measurement and that of the top mass, determined with an experimental uncertainty of 6.5\,MeV ($\sim$30\,MeV, when including theoretical uncertainty) in a $\rm t\bar{t}$ threshold scan, can be compared to the prediction from the ensemble of other electroweak measurements accumulated by FCC-ee, to perform an exacting closure test of the SM, as illustrated in Fig.~\ref{fig:mwvsmt}.

At the FCC-hh, high-energy electroweak processes, such as high-mass Drell-Yan, gauge-boson scattering or associated Higgs production, offer an independent further probe of possible deviations from the SM couplings, arising through the rate enhancement induced by higher-dimension EFT operators.
 
\vspace*{-0.5cm}
\subsection*{Flavour physics}

Precise measurements of heavy-quark transitions can provide sensitivity to new particles at the highest mass scales, and studies of CP violation in beauty and charm decays address a central mystery in nature, the origin of the baryon asymmetry of the universe. In the lepton sector, many BSM theories predict significant effects to manifest themselves in the third generation, making tau physics of great interest.

In heavy-quark studies, FCC-ee will benefit from a pristine analysis environment, not found at the HL-LHC, and samples that are more than ten times larger than those foreseen at Belle~II.  It will also collect the world's largest sample of $\tau$ decays. These datasets will present a wealth of opportunities, including:
\vspace*{-0.2cm}
\begin{itemize}[leftmargin=0.4cm,itemsep=0mm, parsep=0pt]
\item Studies of many very rare b and c hadron decays of high interest that lie far beyond the reach of existing facilities. For example,  the sensitivity in $\rm b \to s \tau^+\tau^-$ transitions will be improved by four orders of magnitude with respect to current limits;
\item Measurements of CP-violating observables in beauty and charm transitions that will match or exceed in precision those expected at LHCb Upgrade~II, particularly in those decays with final-state neutrals, where FCC-ee will have unique capabilities and reach;
\item The ability to search for lepton-flavour violating tau decays down to branching fractions of ${\cal O}(10^{-11})$, and to improve existing tests of lepton universality by at least an order of magnitude. 
\end{itemize}
These, and many other, transformational measurements will only be possible thanks to the extremely high luminosities of the Tera-Z run.  Hence, FCC-ee presents not only outstanding prospects for flavour physics, but  also represents the {\it only} opportunity for these studies to advance significantly at CERN in the post HL-LHC era.  

 Many observables in flavour physics have small or negligible theoretical uncertainties, which motivates striving for measurements of the highest possible precision, even beyond what will be achievable at FCC-ee.
The layout of FCC-hh makes provision for an interaction region that could house a dedicated flavour experiment, as with LHCb at the (HL-)LHC.  This experiment would benefit from the five-times larger $\rm b\bar{b}$ production cross section and higher luminosity of FCC-hh, and the expected advances in detector and computer technology that will improve triggering performance. 

\vspace*{-0.4cm}
\subsection*{Direct searches}
\vspace*{-0.1cm}

\begin{figure}[t]
\begin{center}
\includegraphics[width=0.7\linewidth]{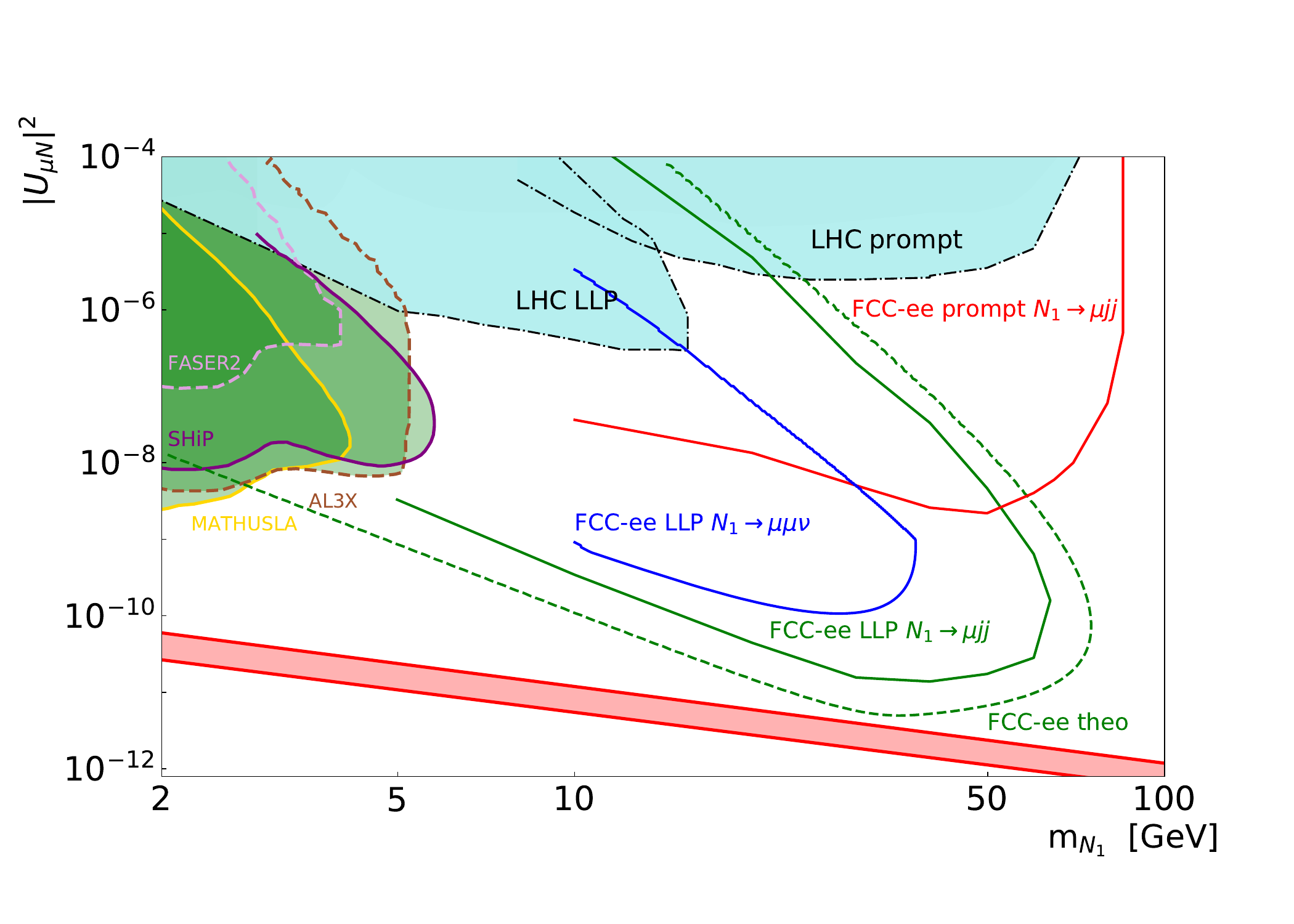}
\end{center}
\vspace*{-0.5cm}
\caption{\small Sensitivity on HNLs in the parameter space of mass against mixing (summed over all flavours). FCC-ee expectations  are shown for two long-lived particle (LLP) searches. The dashed green (`FCC-ee theo') shows the case when three decays are found within the detector from the full Tera-Z run.  The red band indicates the `seesaw' bound.  See Ref.~\cite{FCC-FSR-Vol1} from more details.  
}
\label{fig:HNL}
\end{figure}

A key motivation of the Tera-Z run is to extend the search for feebly interacting particles (FIPs) into a mass range over an order of magnitude above that probed by the dedicated experiments that will operate in the HL-LHC era, and to couplings much smaller than those accessed in (HL-)LHC analyses. Searching over this uncharted expanse of parameter space is a scientific imperative given that the masses and couplings of FIPs are not predicted by theory.  The searches will encompass both prompt signals and those from long-lived particles~\cite{Blondel:2022qqo}, and these signatures are driving factors in the design of FCC-ee detectors.  Heavy neutral leptons (HNLs) are of particular interest, given their possible connection with neutrino masses and the baryon asymmetry.  The FCC-ee will have high sensitivity to HNLs over a wide range of parameters, which will approach the seesaw limit for masses of around 40\,GeV, as shown in Fig.~\ref{fig:HNL}.  Similar exciting prospects exist in the search for axion like particles (ALPs) produced in association with a photon, Z or Higgs boson.  Exotic Higgs decays provide yet another opportunity for new particles to manifest themselves. 

New particle searches are one of the main motivations for FCC-hh, where the high parton centre-of-mass energies will allow for discoveries up to very high masses, seeking signs of Higgs compositeness or Supersymmetry.  To give a few examples, strongly coupled new particles, new gauge bosons ($\rm Z^{'}$ and $\rm W^{'}$) and excited quarks will be discoverable up to masses of 40\,TeV; it will be possible to cover the entire parameter space for certain classes of weakly interacting dark matter; and new Higgs bosons will be observable in the range 5-20\,TeV.

\vspace*{-0.5cm}
\subsection*{QCD and heavy ion physics}


Studies of QCD processes are both important in their own right, and essential for developing the necessary control over many other measurements in the FCC-ee physics programme.  Precise knowledge of $\alpha_S(m_Z^2)$ is necessary for the prediction of cross sections and decay rates; improved perturbative calculations, which must be confronted with data, are required to describe jet dynamics; the best possible knowledge of jet substructure and fragmentation is essential for the training of jet-flavour-taggers, which are vital tools in the Higgs-measurement campaign; finally, the understanding of non-perturbative dynamics (hadronisation, colour reconnection, and final-state interactions) is important for measurements of the $\PW$ mass with hadronic final states and for the analysis of the $\rm t \bar{t}$ threshold scan.  Progress on all of these topics will be possible thanks to the very large event samples available and the high granularity of the FCC-ee detectors, which will be valuable for the investigations of jet substructure.  For example, the combined fit of the relevant Z-resonance pseudo-observables will allow $\alpha_S(m_Z^2)$ to be determined with an experimental precision of $0.1\%$.  These measurements will stimulate the corresponding improvements in theory that will be necessary.

The FCC-hh will allow for a programme of vibrant QCD studies in $\rm pp$ collisions.  For example, assuming that the collider will accommodate $\rm AA$ and $\rm pA$ collisions, the Quark Gluon Plasma phase will start at significantly higher temperatures than occurs at the LHC, will live longer, and extend over a wider spatial region.  More discussion of these opportunities can be found in Ref.~\cite{EPPSU-FCC-QCD}.

\vspace*{-0.5cm}
\subsection*{Non-collider opportunities}

The FCC will be able to support other areas of frontier science apart from collider physics.  Initial ideas for synergetic applications of the injector complex and booster ring of FCC-ee are documented in Ref.~\cite{otherscience1}, and these were revisited in a recent workshop~\cite{otherscience2,FCCee-other-science}.  The high-intensity positron source is of interest for fixed-target dark-matter experiments, surface and material science, plasma acceleration and a positronium Bose-Einstein condensate laser.  There are a wealth of opportunities for producing photon beams at different energies, exploiting the injector, the booster and from Compton-backscattered laser light.  The very low emittance of the booster is of particular interest for its potential to produce a coherent and high intensity source for imaging applications.

\vspace*{-0.3cm}
\section{FCC-ee: enabling characteristics}
\vspace*{-0.1cm}

FCC-ee has several key characteristics that will be central to its success in delivering the physics results summarised above, and indeed providing the opportunity for exceeding the baseline goals.

\vspace*{-0.5cm}
\subsection*{The ultimate machine for precision physics}

The measurement programme at FCC-ee will be characterised by extreme statistical precision. The sample sizes at all energy points will be larger than those achievable by any alternative project proposed at CERN.  This fact is evident when considering the 10.8\,ab$^{-1}$ dataset at 240\,GeV for Higgs studies, which will be collected in only three years.
 Equally striking is the sample size from the Tera-Z phase, which will be $10^5$ greater than that of LEP.
Such statistical sensitivity mandates a correspondingly excellent control over experimental (and theoretical) uncertainties.  A critical attribute of FCC-ee in this respect is the possibility to measure the beam energy through resonant depolarisation, which is a unique property of circular colliders and will be essential for many flagship measurements at the \PZ pole and at the $\PW\PW$ threshold.  The redundancy provided by four experiments, discussed below, is another very powerful feature.

\vspace*{-0.5cm}
\subsection*{The baseline physics programme and beyond}

\begin{figure}[htbp]
\centering
\includegraphics[width=0.95\columnwidth,angle=0]{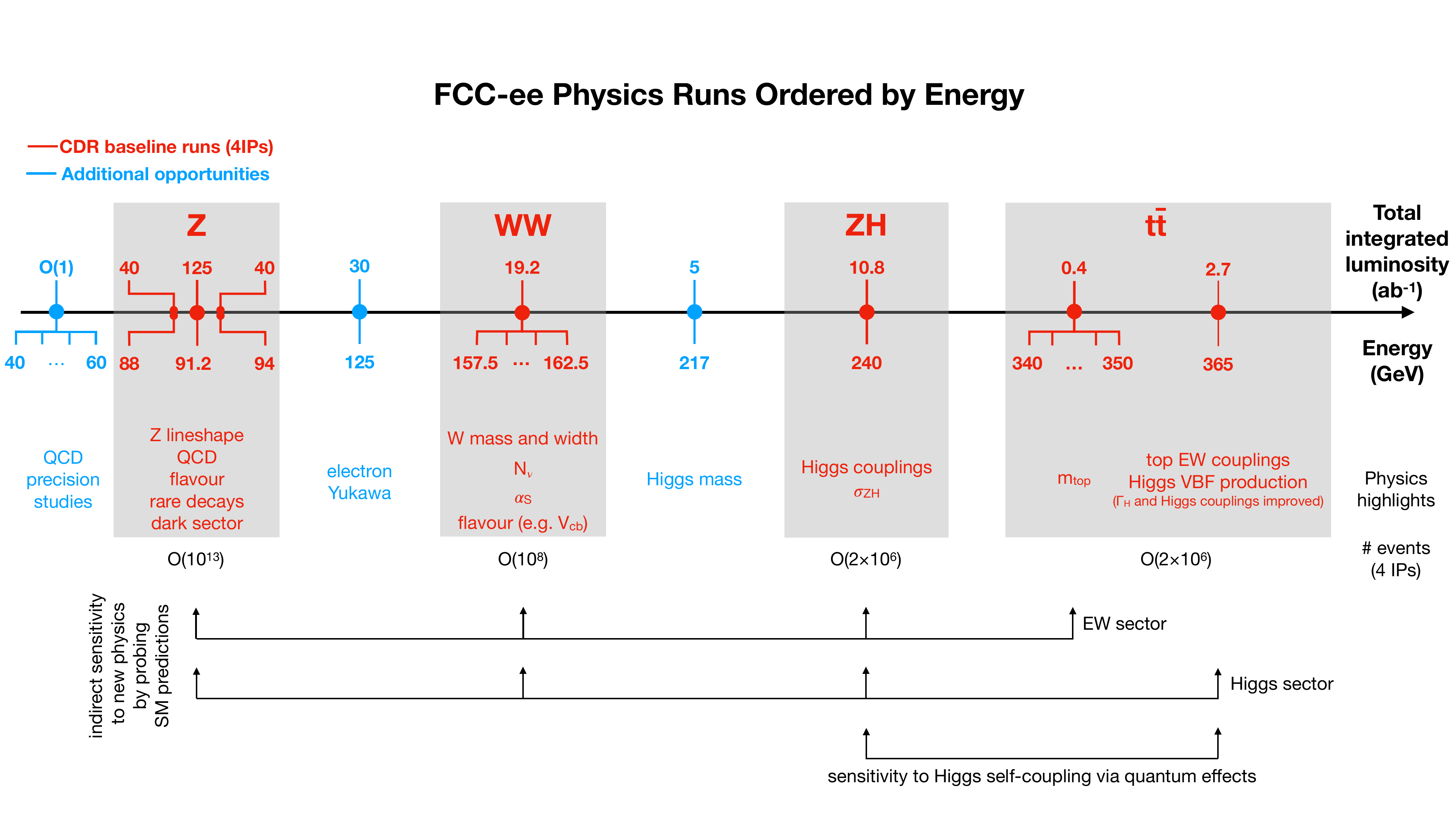}
\caption{\label{fig:cdrplus} 
\small 
Potential physics programme for FCC-ee, ordered by increasing centre-of-mass energy, without indication of a specific chronological sequence. Red indicate the minimal programme with fifteen years of running, and with the corresponding integrated luminosities and physics outcome. The numbers of \PZ, $\PW\PW$, $\PZ\PH$, and $\ttbar$ events delivered to four interaction points are listed.  Blue indicates a possible wider physics programme.}
\end{figure}

The FCC-ee will be a superlative Higgs factory, but its physics programme, summarised above, has enormous breadth, which distinguishes it from alternative projects proposed for CERN.
This wide programme derives from the samples summarised in Table~\ref{tab:seqbaseline} and is illustrated schematically in Fig.~\ref{fig:cdrplus}, which will be collected over a compact period of fifteen years, including one year of shutdown before starting operation at and above the $\PQt\PAQt$ threshold.  This programme corresponds to the current baseline plan for operation; when contemplating this, and looking beyond, several important considerations are to be noted~\cite{note-FCCeeSequence}.

\vspace*{-0.3cm}
\begin{itemize}[leftmargin=0.4cm,itemsep=0.5mm, parsep=0pt]
\item  {\bf A complete programme delivered by a single machine} \,\,  The \PZ-pole, $\PW\PW$ and $\PZ\PH$ datasets will be delivered in less than a decade of continuous operation.  The transition to operation at the highest energies will require only one year of shutdown to prepare the RF system, and amount to no more than 10\% of the total cost of the project.  
%
\item {\bf Flexibility} \,\, There is quasi-total flexibility  in the order of operation for all running points below the $\PQt\PAQt$ threshold.  Although the foreseen datasets are often discussed in order of escalating energy, it would be prudent to defer the bulk of the \PZ pole running to the later years of operation, as the extreme precision aimed for in the Tera-Z programme places the greatest demands on the accelerator, energy calibration and detectors.  However, there is no need to decide on the precise order of data taking at this early stage of planning -- FCC-ee allows for great agility in shaping and adjusting the schedule of operation.
\item {\bf Extendability} \, \, The period of operation at each energy point and foreseen sample sizes in Table~\ref{tab:seqbaseline} should be regarded as a baseline proposal, which balances the physics goals of the project in a manner commensurate with current knowledge and assumed priorities.   It is important to recall that a large fraction of physics measurements at FCC-ee will be statistics limited, e.g. many EW observables such as $\rm \alpha_{QED}(\rm m_Z)$ and the W mass, the flavour observables, BSM searches, the measurement of the Higgs couplings and loop-level determination of the Higgs self coupling, and so the programme may evolve to reflect a new optimisation of these goals.  
Furthermore, intriguing results, either from the HL-LHC or from early running, may motivate additional data-taking at a given working  point.
Any extension in the duration of FCC-ee operation, whether motivated by physics or external factors, can be fully exploited.
Indeed, it will be possible to enrich the baseline programme, through the inclusion of additional energy points, as indicated in Fig.~\ref{fig:cdrplus}.  Possibilities include operating at the Higgs pole (125\,GeV) in order to access the electron-Higgs coupling, running at around 217\,GeV in order to measure the Higgs mass through its threshold cross-section, and 
 collecting samples of a few ab$^{-1}$ at centre-of-mass energies of 40 and 60\,GeV for precision QCD studies.  All of these prospects are under active evaluation, with the sensitivity of the Higgs-pole run  being dependent on the development of an effective monochromatisation scheme~\cite{FCC-FSR-Vol1,FCC-FSR-Vol2}.
\end{itemize}
\vspace*{-0.7cm}
\subsection*{Four interaction points}

A highly advantageous attribute of a high-energy synchrotron is its ability to supply collisions to multiple interaction points (IPs) at a luminosity similar to that which a single IP would receive.  An important recommendation from the mid-term review of the Feasibility Study was to increase the number of IPs from two to four.  This is advantageous for many reasons, which include:
\vspace*{-0.2cm}
\begin{itemize}[leftmargin=0.4cm,itemsep=0.5mm, parsep=0pt]

\item {\bf Detector redundancy and diversity} \, \, The extremely broad physics programme of FCC-ee requires a diversity of detector solutions to ensure full coverage of the science goals, e.g. detectors optimised for flavour physics or long-lived particle searches, as well as Higgs and electroweak physics, and to provide redundancy in performance. This requirement can be satisfied with four IPs, but not with two.
\item {\bf Robustness and protection against unforeseen systematic biases} \,\, The experience of LEP provides many examples of how four IPs enabled biases present in one experiment to be revealed through a comparison with the results of the others.  A striking example is the discovery in 1991 data of an offset of around 20\,MeV between the measurement of the \PZ mass in OPAL and L3 compared with that in ALEPH and DELPHI, which was then understood to arise from overlooking the evolution in beam energy around the ring through synchrotron radiation and RF boosts. This bias would have risked remaining undiscovered on an accelerator with only two experiments.
\item {\bf Increase in integrated luminosity}\,\, The evolution from the minimal design of two to four IPs increases the total integrated luminosity that will be collected by a factor of 1.7.  This brings great benefits, whether viewed in terms of increased precision and discovery potential, or shorter running time required to attain the initially defined physics goals.
\item {\bf Community building} \,\, 
The challenges of the FCC-ee will call for creative partnerships across the entire HEP community, which can be better accommodated in four experiments rather than two, with each collaboration providing an environment for individual ingenuity to flourish.
Four collaborations at FCC-ee will also serve as excellent academies in which to train and grow the community in preparation for the challenges of FCC-hh.
\end{itemize}
\vspace*{-0.7cm}
\subsection*{High efficiency} 

FCC-ee is the project that will deliver the highest priority physics goals of the HEP community with the highest efficiency, when viewed by any metric.  This can be most clearly seen by focusing on the Higgs programme alone, which within the baseline plan of FCC-ee is completed within nine years, including one year of shutdown.   The very high luminosity of the collider and four IPs means that the sensitivities attained on the Higgs couplings significantly exceed those of other proposed projects, as demonstrated in Ref.~\cite{Blondel:2024mry}.  Viewed alternatively, the programme time at other facilities would likely have to be much longer than that at FCC-ee in order to reach similar precision, leading to correspondingly greater energy usage, carbon footprint and  operational costs, as well as the associated problem of maintaining community interest on a relatively narrow programme  over a period of several decades.

\vspace*{-0.3cm}
\section{FCC-hh and strengths of the integrated programme}
\vspace*{-0.1cm}

A high energy hadron collider is the natural and necessary second step of the FCC integrated programme. The FCC-hh possesses many strengths, beyond its key attributes as a search machine and high-energy Higgs factory, as does the integrated programme as a whole.


\vspace*{-0.4cm}
\subsection*{Physics robustness against choice of running energy and luminosity}

The baseline FCC-hh configuration presented in the FSR~\cite{FCC-FSR-Vol2} foresees the use of 14\,T Nb$_3$Sn dipoles, rather than  the 16\,T magnets assumed in the CDR~\cite{FCC-hhCDR}.  This evolution in design, and the final choice of tunnel circumference (90.7\,km) leads to a collision energy of 84.6\,TeV, reduced with respect to the CDR, but with a similar ultimate luminosity of $30 \times 10^{34}$\,cm$^{-2}$s$^{-1}$.  
Since the FSR, studies have explored the dependence in physics sensitivity when considering a range of collision energies from 72\,TeV, which is a conceivable scenario in the event of an accelerated schedule and the deployment of 12\,T dipoles, to 120\,TeV, which may be reachable with advances in high-temperature superconducting technology~\cite{EPPSU-FCC-FCChh}. Assuming a fixed power budget for cryogenics, the running luminosity would fall with energy on account of the increased synchrotron radiation.  
Reference~\cite{EPPSU-FCC-FCChh} presents examples of the interplay between energy and luminosity in defining the physics performance for both precision measurements and searches at the high-mass end. These preliminary studies demonstrate the robustness of the key physics targets against small variations of the machine parameters around the baseline configuration.
It is noted, however, that the physics potential  of a machine of significantly lower collision energy, such as the so-called High Energy LHC operating at $\sqrt{s}=36$\,TeV, is inferior by factors of 2-3 in mass reach and by a factor of 2 in precision on Higgs couplings~\cite{LE-FCCphysics}.

\vspace*{-0.5cm}
\subsection*{Complementarities and synergies}

The two phases of the FCC integrated programme play equally important and complementary roles in the characterisation of the Higgs sector.  At FCC-ee, subpercent precision will be attained on many of the most interesting coupling parameters, and several couplings will be probed that are inaccessible at hadron machines, e.g. $\rm H c\bar{c}$,   $\rm H s\bar{s}$, and possibly $\rm H e^+e^-$.  At FCC-hh precise measurements will be made of the most suppressed Higgs couplings, e.g. $\rm H \gamma\gamma$, $\rm H\mu^+\mu^-$ and $\rm HZ\gamma$. Most importantly, a deep understanding will be attained of the Higgs potential, with a percent-level determination of the Higgs self coupling.   The success of these studies are contingent on the synergies that exist across the programme.  The measurements of FCC-ee at the Z pole will allow electroweak systematics to be eliminated from the interpretation of the Higgs observables.  The measurement of the $\rm HZZ$ coupling and the Higgs width at FCC-ee will permit all the branching fractions at both machines to be determined in a model-independent manner. Measurements of the $\rm Zt\bar{t}$ coupling at FCC-ee can be used to normalise the $\rm t\bar{t}H$ cross-section at FCC-hh to the $\rm t\bar{t}Z$ cross-section, and enable a $\sim$1\% determination of the top-Yukawa coupling.

These complementarities and synergies are seen in most other physics areas of the FCC.  Deviations from SM predictions in precision measurements at FCC-ee will lead to the expectation of new massive particles, which can then be searched for at FCC-hh.   The direct search-capabilities of both machines is complementary, with FCC-ee being highly sensitive to HNLs, ALPs and other FIPs for masses below $\sim$100\,GeV, while FCC-hh can discover higher mass new particles across a very broad range of scenarios, for example $\rm Z^{'}$ states up to masses of $\sim$40\,TeV. 
\vspace*{-0.5cm}
\subsection*{Physics diversity}

It follows from the enormous scope of the physics output from the LHC that FCC-hh will have a similarly broad science programme, beyond that conceivable at any alternative energy-frontier machine.  
Full exploitation of the collider opportunities 
 will almost certainly require dedicated experiments (e.g. focused on flavour and heavy-ion studies) in addition to the general purpose detectors that target high-$\rm p_T$ physics.   In addition, there will be exciting possibilities for performing measurements with far-forward detectors, which have been pioneered at the LHC~\cite{FASER:2024hoe,SNDLHC:2023pun}, where a proposal exist for a Forward Physics Facility~\cite{Feng:2022inv}.  Such detectors would have even richer physics opportunities at the FCC-hh, both in BSM searches and in harnessing the enormous samples of multi-TeV neutrinos that will be produced~\cite{MammenAbraham:2024gun}. Around $10^9$ electron/muon and $10^7$ tau neutrinos could be detected, which would be invaluable for DIS studies of nucleon and nuclear structure, and making measurements relevant for particle astrophysics.

\vspace*{-0.5cm}
\subsection*{Extendability and the lasting benefits of a common infrastructure}

The FCC integrated programme, as conceived today, will enable frontier-physics exploration for a period of over fifty years (including the decade needed to transition from FCC-ee to FCC-hh).  The common infrastructure required by the two machines, most notably the tunnel, is therefore seen to be a remarkably efficient investment, both in terms of capital cost and in the carbon budget from the civil engineering.

Extensions to the baseline programme can already be envisaged.  As discussed above, the rich physics opportunities of FCC-ee are far from saturated by the currently foreseen fifteen years of operation.  Similar considerations apply to FCC-hh, for example in the domain of $\rm ep$ collisions.   The physics opportunities from such collisions have been studied in the context of the LHC, most promisingly when enabled through the construction of an energy-recovery linac~\cite{LHeCStudyGroup:2012zhm}.  This configuration has also been explored for the FCC-hh~\cite{FCC-hhCDR}, where centre-of-mass energies of over 3\,TeV would be achievable, providing interesting opportunities in Higgs measurements, electroweak physics, QCD studies and BSM searches.  The FCC infrastructure could also be repurposed to house a fast accelerator and injector for a very high energy muon collider in the LEP/LHC tunnel.

\vspace*{-0.5cm}
\section{Conclusion}
\vspace*{-0.1cm}

At the completion of the HL-LHC, the LEP-LHC integrated programme will have delivered sustained scientific excellence for over 50 years, and greatly advanced our understanding of the fundamental interactions.   The succession of FCC-ee and FCC-hh
in a common tunnel will replicate and magnify this success, with vastly better precision and increased energies. The FCC integrated programme offers outstanding prospects for progress in a wealth of topics, including Higgs, electroweak, and flavour physics, as well as remarkable sensitivity in direct searches for both feebly coupled low-mass particles and those that may exist up to many tens of TeV. 
The current ESPP Update provides an opportunity for the global HEP community to endorse this vision, and then to work together so as to enable the first stage of this project, FCC-ee, to be realised in a timely fashion. Seizing this opportunity will be an important step forward in the journey towards a more complete understanding of the laws of nature.  

\newpage
\appendix
\section{Organisation of FCC physics, experiments and detectors activities, and documentation}
\label{sec:organisation}

 This appendix provides brief information on the organisation of the Physics, Experiments and Detectors (PED) pillar of the FCC project, which has produced the results presented in Volume~1 of the FSR~\cite{FCC-FSR-Vol1} and is preparing for the next phase of the studies and preparatory work.

\begin{figure}[htb]
    \centering
    \includegraphics[width=1.1\linewidth]{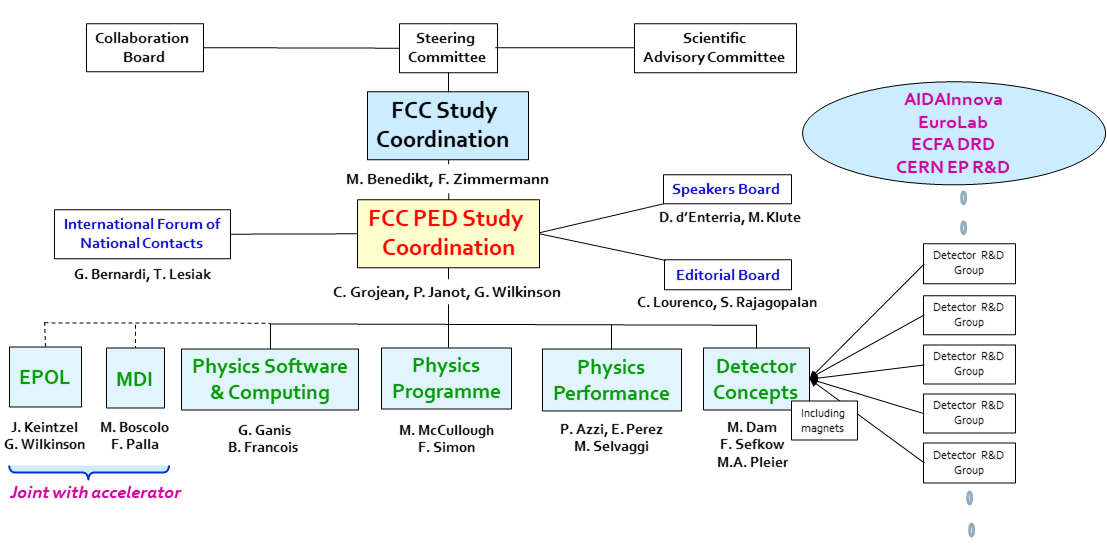}
    \caption{Organisational structure of the Physics, Experiments and Detectors pillar, also showing relevant external bodies.  The names of the current coordinators, conveners, chairs etc. are indicated.}
    \label{fig:PED_organo_overall}
\end{figure}

\begin{figure}[htb]
    \centering
    \includegraphics[width=1.1\linewidth]{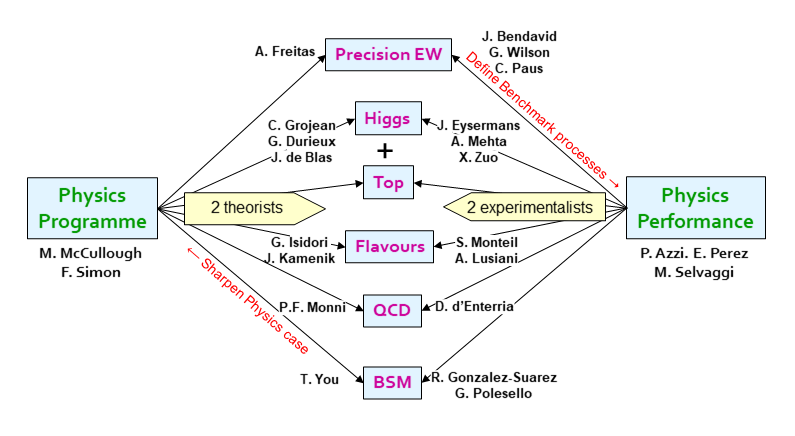}
    \caption{Organisational structure of the physics groups, showing the relationship to the Physics Programme and Physics Performance WPs. The names of the current coordinators, conveners, chairs etc. are indicated.}
    \label{fig:PED_organo_physics}
\end{figure}

 Figure~\ref{fig:PED_organo_overall} shows a schematic of the current PED organisational structure, and how it relates to other bodies within the FCC project as a whole, and external initiatives. 
The {\bf PED Study Coordination} oversees six Work Packages (WPs):  
 \begin{itemize}
 \item The {\bf EPOL} WP, which is concerned with the calibration of the collision energy and monochromatisation studies, and the {\bf Machine Detector Interface (MDI)} are joint WPs also under the Accelerator pillar.  
 \item The {\bf Physics Programme} WP maps out the science opportunities of the FCC programme and ensures the necessary generators are available, and calculations performed.  The {\bf Physics Performance} WP investigates how this programme can be implemented, in particular simulating certain measurements that are useful case studies for defining detector requirements and shows what physics reach can be expected, given a certain detector performance.  There are six physics groups common to both these WPs, concerned with the main physics topics of interest:  precision electroweak, Higgs physics, top physics, flavour physics, QCD and BSM physics (see Fig.~\ref{fig:PED_organo_physics}).  
 \item The {\bf Detector Concepts} WP receives the requirements from the Physics Programme WP and develops and simulates (sub-)detector designs that aim to meet these requirements, and feeds back information on the performance of these designs.  It interacts with external bodies that are performing R\&D into detector technology, in particular the recently formed DRD collaborations.
 \item The {\bf Physics Software and Computing} WP formulates a strategy for software and computing for the experiments during the FCC era, as well as providing support for the current needs of the other WPs.
\end{itemize}

The {\bf International Forum of National Contacts (IFNC)} facilitates the expansion and strengthening of the collaboration of FCC physicists by attracting additional HEP groups, and by helping the structuring of national communities to put them in a better position to obtain resources. A quasi-complete list of European states ($\sim$25) is now represented in the IFNC, together with many other nations, including Argentina, Brazil, Canada, Chile, India, Mexico, Pakistan, South Korea, Turkey and the USA.

In addition, there is a {\bf Speakers Board} and {\bf Editorial Board}.

If the outcome of the 2025-2026 ESPP Update is favourable for the FCC, it will be necessary to broaden the level of involvement of the international community in all PED activities. The Detector Concepts WG is already facilitating the transition to this next phase, and has recently issued a call for expressions of interest into detector concepts and sub-detector systems~\cite{EPPSU-FCC-detconcepts}.
The PED coordinators and chairs of the IFNC welcome discussions with any potentially interested parties, who may also join the relevant egroups, information on which is given in Fig.~\ref{fig:PEDemail}.  

\newcommand\RedEnvelop[1]{\href{mailto:#1}{\includegraphics[width=.4cm]{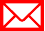}}}
\newcommand\GreenEnvelop[1]{\href{mailto:#1}{\includegraphics[width=.4cm]{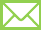}}}
\newcommand\GreenPeople[1]{\href{#1}{\includegraphics[width=.4cm]{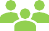}}}

\begin{figure}[htb]
    \centering
    \begin{tcolorbox}
\begin{itemize}
\item[$\circ$] {\bf PED Feasibility Study} \RedEnvelop{FCC-PED-FeasibilityStudy@cern.ch}
	\subitem \hspace{-.5cm}$>$ Physics Programme  \RedEnvelop{FCC-PED-PhysicsProgramme@cern.ch} and Physics Performance  \RedEnvelop{FCC-PED-PhysicsPerformance@cern.ch}
		\subsubitem $\dashv$ QCD \GreenEnvelop{FCC-PED-PhysicsGroup-QCD@cern.ch} \GreenPeople{https://groups-portal.web.cern.ch/group/fcc-ped-physicsgroup-qcd/details}
		\subsubitem $\dashv$ Higgs \GreenEnvelop{FCC-PED-PhysicsGroup-Higgs@cern.ch} \GreenPeople{https://groups-portal.web.cern.ch/group/fcc-ped-physicsgroup-higgs/details}
		\subsubitem $\dashv$ EW \GreenEnvelop{FCC-PED-PhysicsGroup-EWPrecision@cern.ch} \GreenPeople{https://groups-portal.web.cern.ch/group/fcc-ped-physicsgroup-ewprecision/details}
		\subsubitem $\dashv$ Top \GreenEnvelop{FCC-PED-PhysicsGroup-Top@cern.ch} \GreenPeople{https://groups-portal.web.cern.ch/group/fcc-ped-physicsgroup-top/details}
		\subsubitem $\dashv$ Flavour \GreenEnvelop{FCC-PED-PhysicsGroup-Flavours@cern.ch} \GreenPeople{https://groups-portal.web.cern.ch/group/fcc-ped-physicsgroup-flavours/details}
		\subsubitem $\dashv$ BSM \GreenEnvelop{FCC-PED-PhysicsGroup-BSM@cern.ch} \GreenPeople{https://groups-portal.web.cern.ch/group/fcc-ped-physicsgroup-bsm/details}
	\subitem \hspace{-.5cm}$>$ Hadron Collider Studies \RedEnvelop{FCC-PED-HadronColliderStudies@cern.ch}
		\subsubitem $\dashv$ ESPPU25 \GreenEnvelop{FCC-PED-hh-ESPP25@cern.ch} \GreenPeople{https://groups-portal.web.cern.ch/group/fcc-ped-hh-espp25/details}
		\subsubitem $\dashv$ Physics Performance \GreenEnvelop{FCC-PED-hh-PhysicsPerformance-ESPP25@cern.ch} \GreenPeople{https://groups-portal.web.cern.ch/group/fcc-ped-hh-physicsperformance-espp25/details}
	\subitem \hspace{-.5cm}$>$ Detector Concepts \GreenEnvelop{FCC-PED-DetectorConcepts@cern.ch} \GreenPeople{https://groups-portal.web.cern.ch/group/fcc-ped-detectorconcepts/details}
	\subitem \hspace{-.5cm}$>$ Software and Computing \RedEnvelop{FCC-PED-SoftwareAndComputing@cern.ch}
		\subsubitem $\dashv$ Developers \GreenEnvelop{FCC-PED-SoftwareAndComputing-Developers@cern.ch} \GreenPeople{https://groups-portal.web.cern.ch/group/fcc-ped-softwareandcomputing-developers/details}
		\subsubitem $\dashv$ Generators \GreenEnvelop{FCC-PED-SoftwareAndComputing-Generators@cern.ch} \GreenPeople{https://groups-portal.web.cern.ch/group/fcc-ped-softwareandcomputing-generators/details}
		\subsubitem $\dashv$ Full Simulations \GreenEnvelop{FCC-PED-SoftwareAndComputing-Full-Simulation@cern.ch} \GreenPeople{https://groups-portal.web.cern.ch/group/fcc-ped-softwareandcomputing-full-simulation/details}
		\subsubitem $\dashv$ Analysis \GreenEnvelop{FCC-PED-SoftwareAndComputing-Analysis@cern.ch} \GreenPeople{https://groups-portal.web.cern.ch/group/fcc-ped-softwareandcomputing-analysis/details}
		\subsubitem $\dashv$ Detector \GreenEnvelop{CC-PED-SoftwareAndComputing-Detector@cern.ch} \GreenPeople{https://groups-portal.web.cern.ch/group/fcc-ped-softwareandcomputing-detector/details}
		\subsubitem $\dashv$ MDI \GreenEnvelop{FCC-PED-SoftwareAndComputing-MDI@cern.ch} \GreenPeople{https://groups-portal.web.cern.ch/group/fcc-ped-softwareandcomputing-mdi/details}
		\subsubitem $\dashv$ Resources \GreenEnvelop{FCC-PED-SoftwareAndComputing-Resource@cern.ch} \GreenPeople{https://groups-portal.web.cern.ch/group/fcc-ped-softwareandcomputing-resource/details}
		\subsubitem $\dashv$ Documentation \GreenEnvelop{FCC-PED-SoftwareAndComputing-Documentation@cern.ch} \GreenPeople{https://groups-portal.web.cern.ch/group/fcc-ped-softwareandcomputing-documentation/details}
	\subitem \hspace{-.5cm}$>$ Machine Detector Interface \GreenEnvelop{FCC-ee-MDI@cern.ch} \GreenPeople{https://groups-portal.web.cern.ch/group/fcc-ee-mdi/details}
	\subitem \hspace{-.5cm}$>$ Polarisation and Energy Calibration \GreenEnvelop{FCC-ee-PolarizationAndEnergyCalibration@cern.ch} \GreenPeople{https://groups-portal.web.cern.ch/group/fcc-ee-polarizationandenergycalibration/details}
\end{itemize}
\end{tcolorbox}
    \caption{Organisation of FCC PED email groups, where the green indicates that subscription is possible while the red {\tt e-groups} are cumulative groups built from the open green groups underneath (and therefore no direct subscription can be made).  The people icon leads to an information page, and the envelope icon sends a mail (digital version only). }
    \label{fig:PEDemail}
\end{figure}

This \href{https://repository.cern/communities/fcc-ped-sub/records?q=&l=list&p=1&s=10&sort=newest}{link} provides access to a repository of FCC PED notes, documenting studies and recent progress.   A collection of essays on the the challenges and opportunities of the FCC physics programme, published in Eur. Phys. J. Plus may be found \href{https://link.springer.com/journal/13360/topicalCollection/AC_e20d0ca1d36bc88d0e8c796d3f2e083a}{here}.


\cleardoublepage
\bibliographystyle{cms_unsrt}
\bibliography{biblio}%

\end{document}